\documentclass[]{aastex631}
\usepackage{wrapfig}

\accepted{December 15, 2022}

\submitjournal{ApJ}

\shorttitle{Equilibrium of Loops Near Separatrices}
\shortauthors{Mason et al.}

\begin{document}

\title{The Equilibrium of Coronal Loops Near Separatrices}

\correspondingauthor{Emily I Mason}
\email{emason@predsci.com}

\author[0000-0002-8767-7182]{Emily I Mason}
\affiliation{Predictive Science Inc. \\
9990 Mesa Rim Rd, Suite 170 \\
San Diego, CA 92121, USA}

\author[0000-0003-0176-4312]{Spiro K Antiochos}
\affiliation{University of Michigan \\
500 S. State Street  \\
Ann Arbor, MI 48109, USA}

\author{Stephen Bradshaw}
\affiliation{Rice University \\
6100 Main St.  \\
Houston, TX 77005-1827, USA}

\begin{abstract}
We present numerical models from the field-aligned Hydrodynamics and Radiation Code (HYDRAD) of a highly asymmetric closed coronal loop with near-singular expansion factor. This loop was chosen to simulate a coronal magnetic flux tube that passes close to a null point, as in the last set of closed loops under the fan surface of a coronal jet or a pseudostreamer. The loop has a very large cross-section localized near the coronal null. The coronal heating was assumed to be uniform and steady. A siphon flow establishes itself within 4 hours of simulation time, flowing from the smaller-area footpoint to the larger-area footpoint, with high initial speeds dropping rapidly as the plasma approaches the null region. Observationally, this would translate to strong upflows on the order of 10 km s$^{-1}$ from the footpoint rooted in the localized minority polarity, and weak downflows from the fan-surface footpoint on the order of a few km s$^{-1}$, along with near stationary plasma near the null region. We present the model results for two heating rates. In addition, we analyzed analogous Hinode EIS observations of null-point topologies, which show associated Doppler shifts in the plasma that correlate well with the simulation results in both direction and magnitude of the bulk velocity. We discuss the implications of our results for determining observationally the topology of the coronal magnetic field.
\end{abstract}


\section{Introduction} \label{sec:intro}

The structure and dynamics of the solar atmosphere are widely-believed to be determined by two major effects: the magnetic field, which is due primarily to the sources at the photosphere, and the coronal heating process that determines the properties of the plasma contained in this field. Since the plasma beta is measured to be low in the corona ($\beta << 1$), the field magnitude is fairly uniform \citep{Schrijver1999}. The coronal heating, on the other hand, must have considerable structure because it can only occur at small scales for the high-Lundquist numbers of the corona. This in turn structures the plasma, but only across field lines, because both mass transport and thermal transport are highly efficient along the field. The net result is that coronal plasma is distributed with roughly uniform temperature and density along the field, but with clear variations across, so the corona appears as a system of loops.

Observational analysis and numerical modeling of coronal loops have been active fields of study for decades, since shortly after the first X-ray observations by Skylab showed (apparently) monolithic tubes of hot plasma in the low corona (e.g., \cite{Rosner1978}). These studies found that the loop plasma achieves a quasi-steady equilibrium described fairly well by the celebrated scaling laws \citep{Craig1978,Rosner1978,Vesecky1979}, which predict that the loop plasma temperature and density are approximately constant in the corona, and determined by only the heating rate and the loop length, L. The expansion factor, \textit{$\Gamma$}, measures the cross-sectional variation of the loop from its chromospheric base to its apex; this plays a more minor role on the plasma properties  \citep{Vesecky1979,Cargill2021}. For the most part plasma velocities are small, $< 10 \, km \, s^{-1}$, due to asymmetries introduced by spatial variations of the heating or by asymmetries in the loop geometry, i.e., spatial variations in \textit{$\Gamma$}.  Strong dynamics in coronal loops are possible (i.e., if the heating is strongly episodic as in certain nanoflare models \citet{Parker1988,Klimchuk2006}, or strongly localized as in the case of thermal non-equilibrium \citet{Antiochos1991,Mason2019}). However, sophisticated numerical models with a self-consistent treatment for wave heating show, for the most part, agreement with the quasi-static scaling-law loops \citep{Oran2013,Lionello2009,Torok2018}. 

\begin{wrapfigure}[39]{l}{0.5\textwidth}
\vspace{-\intextsep}
\centering
    \hspace*{-.75\columnsep}
    \includegraphics[width=1.\linewidth,clip]{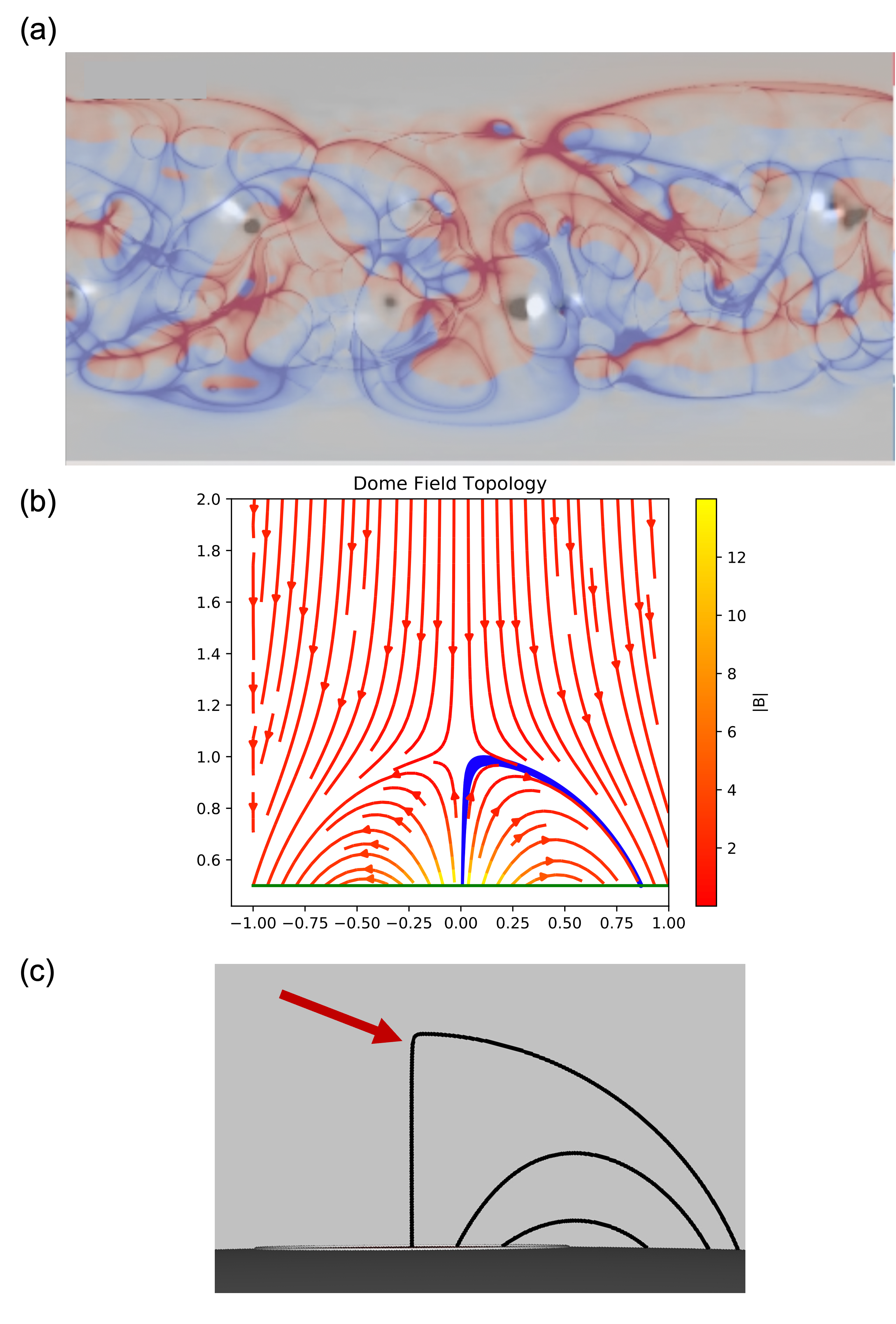}
              \caption{(a) Signed $log(Q)$ (red and blue S-web map) over an example HMI magnetogram. (b) diagram showing overall magnetic fields of a null-point topology, with example flux tube outlined in blue. Credit: Peter MacNeice, private communication. (c) diagram of selected field lines crossing a closed polarity inversion line separating strong minority field from weak surrounding field. The red arrow indicates the field line used to extrapolate flux tube structure. Credit: Richard DeVore, private communication.}
              \label{F1}
\end{wrapfigure}
In recent years, it has been realized that the field in the corona must have topological boundaries, since the magnetic flux distribution at the photosphere is generally complex, exhibiting multiple polarity regions separated by polarity inversion lines (PILs). An example of such a flux distribution can be seen in Figure \ref{F1}a, which shows the smoothed photospheric radial field for Carrington Rotation 2099 in 2010 as measured by HMI on SDO, (see \url{hmi.stanford.edu/QMap/}). The field strength is given by white (positive) and black (negative). We note the presence of several active regions, appearing as bright black and white bipoles, during this CR. It should be emphasized that there are very many more bipoles on the surface, the so-called magnetic carpet, but these are small-scale and are averaged out in this smoothed magnetogram.

Even the large flux concentrations, however, produce considerable structure at the photosphere, which gives rise to a coronal magnetic field with complex topology. For the corona, the topology is given primarily by the connectivity of the field, the connections between the various flux systems. In general, the coronal field is not observed to have higher-order topological features such as knots and disconnected rings. Figure \ref{F1}a also shows the connectivity of the magnetic field as given by the so-called squashing factor, Q \citep{Titov07}. Q is a measure of the gradients in the mapping defined by the magnetic field lines. In the particular case of Figure \ref{F1}a, the mapping is from the spherical surface at 1.1 R$_S$ down to the photosphere, and the contours shown are for large values of $log(Q)$, colored by the polarity of the field, red (negative) and blue (positive). The intricate web of strong Q contours shown in the figure is referred to as the S-Web \citep{Antiochos2011a}. This S-Web indicates locations where the magnetic connectivity is discontinuous, as at a flux boundary, or near-discontinuous, as in the case of so-called quasi-separatrix layers, \citep[e.g.,][]{Demoulin96}.   

There are two key points to the S-Web shown in Fig. \ref{F1}a. First, wherever there is an arc of large Q, electric currents can form easily, leading to magnetic reconnection and eruptive solar activity. This result is the physical underpinning for essentially all models of coronal jets and flares \citep{Shibata1994,Demoulin96,Antiochos99,Wyper2018}. The second key point is that unlike the actual magnetic flux at the photosphere, the Q-map shown in the figure is derived from a model -- in particular, a potential field source surface model. Given that the coronal field is never potential, (there are always filament channels and sheared/twisted structures present), we conclude that the locations of all the arcs and lines in Fig. \ref{F1}a are likely to be incorrect. It would be extremely valuable, therefore, to have some  method for determining observationally the locations of the separatrix curves as in Fig. \ref{F1}a. We propose a possible method in this work.

The S-Web of Fig. \ref{F1}a consists of a variety of separatrix and quasi-separatrix curves, but the simplest such curve (and generally the most common) is that due to a single polarity region surrounded by an opposite-polarity background, giving rise to the so-called embedded bipole topology as seen in Figure \ref{F1}b \citep{Antiochos1998a,Pariat2009}. This topology consists of a dome-like separatrix surface in the corona, the fan surface that separates the flux of the embedded bipole from surrounding flux. Somewhere on the fan the field vanishes to form a 3D null point, and two singular field lines -- the “spines” -- emanate from the null \citep[e.g.,][]{Shibata1994,Young2014,Karpen2017,Wyper2018}. Such a magnetic topology occurs for a single polarity region that has fairly circular geometry, but the actual photosphere contains many polarity regions, sometimes with nested embedded polarities and/or with extremely structured shapes. As a result, the magnetic topology in the corona consists of a complex web of interconnected separatrices and numerous null points/spines, as implied by the figure.

The S-Web has been postulated to have major effects on the structure and dynamics of the solar wind \citep{Antiochos2011a,Titov2011,Linker2011,Higginson2017}, especially the slow wind, but has generally not been considered in modeling the closed field plasma. The primary reason for this is that the topology of the coronal field is very difficult to observe directly in XUV or X-ray images. Only certain topological features, such as domes and spines have been observed, and only on the limb \citep[e.g.,][]{Mason2021}. Note also that the S-Web and topological features are generally global quantities and cannot be determined from local measurements of the magnetic field, even if such were available. Consequently, the existence of the S-web itself is currently only inferred from theoretical models.  

In this paper we propose using loop models to determine observational proxies of coronal topology, specifically, the connectivity of the magnetic field.   The key point for our studies is that flux tubes near the features of the S-Web must exhibit extreme variations in their geometry (such as that indicated by the highlighted field line in Figure \ref{F1}c), because they must pass close to a null point, and consequently, will have near-singular expansion factor \textit{$\Gamma$} there. Such variations are likely to produce observable signatures in the coronal plasma, such as rapid flows from one end to the other. By determining the characteristics expected for these signatures, it should be possible to map out the S-Web using high-resolution spectroscopic observations expected from missions such as the upcoming EUVST \citep{Shimizu2019} and MUSE \citep{DePontieu2022}, for example.

Furthermore, investigating flux tubes near topological singularities may well result into new insights into coronal loops, themselves. It is not clear that equilibrium states are even possible for extreme cross-section variation, even if the heating is spatially and temporally constant. It is well known that extreme localization of the heating process can lead to a lack of equilibrium, a so-called thermal non-equilibrium \citep{Antiochos1991}; the same may occur if the expansion is extremely localized.

For this first study of the properties of S-Web plasma, we adopt the classic assumption of a uniform and constant coronal heating and consider the most common near-singular magnetic structure, a flux tube that lies close to the separatrix surface of an embedded bipole and, therefore, passes very near to a coronal null. Such flux tubes must undergo a dramatic, localized increase in cross-sectional area, in order to satisfy the conservation of magnetic flux. These flux tubes will be asymmetric in two ways: the aforementioned varying cross-sectional area, and a left-to-right asymmetry due to the fact that one leg of the flux tube lies near the inner spine (associated with stronger field) whereas the other leg lies near the (weaker-field) fan surface \citep{Mason2019}. This results in a strong global area factor variation in addition to the localized maximum near the null. This type of coronal loop, modeled here, can be seen in blue in Figure \ref{F1}b.

In the following section we describe the field-aligned hydrodynamic code, HYDRAD, that we use and provide the details of the loop model: geometry, heating rate, etc. Section \ref{sec:results} presents the results of the simulation, including an analysis of expected observational signatures. For comparison with these results, the next section shows Hinode EUV Imaging Spectrometer (EIS, \cite{Kosugi2007,Culhane2007}) observations of similar loops in several medium-sized embedded bipoles like those discussed above, called null-point topologies (NPTs) near disk center. The final section discusses the implications of this study for observational determination of the S-web in closed field regions, and for the fundamental physics of coronal loops. Furthermore, we discuss the likely effects of important extensions to the model, such as a spatially or temporally varying heating and effects of magnetic field dynamics.

\section{HYDRAD}\label{sec:hydrad}

\subsection{Equations}

For the simulations presented here, we use HYDRAD \citep{Bradshaw2003,Bradshaw2013}, a field-aligned, adaptive mesh, hydrodynamic code that can accommodate arbitrary loop geometries via a polynomial fit to the cross-sectional area factor and parallel component of gravity along the field. HYDRAD is well-documented and has been widely used for calculating coronal loop plasma dynamics in a wide variety of situations, including nanoflare heating and thermal nonequilibrium \citep{Bradshaw2004,Bradshaw2005,Bradshaw2011a,Cargill2012,Brooks2016,Reep2016,Johnston2019}. We do not utilize the partial ionization state capabilities of the code, since these generally are more relevant in impulsive heating scenarios. The loops we model here have only steady background heating, rendering the special ionization rates less likely to be necessary. It should be noted, however, that a fast steady flow through the transition region can affect the radiative losses in that region, but the qualitative structure remains the same (e.g., \cite{Spadaro1990}).

HYDRAD solves the conservative form of the standard hydrodynamic loop equations, excerpted partially here from \citet{Bradshaw2003,Bradshaw2013}: 

\begin{equation}
    \frac{\partial \rho}{\partial t}+\frac{1}{A(s)}\frac{\partial}{\partial s}(A(s)\rho v)=0
\end{equation}

\begin{equation}
    \frac{\partial}{\partial t}(\rho v)+\frac{1}{A(s)}\frac{\partial}{\partial s}(A(s)\rho v^2)=\rho g_{||} - \frac{\partial P}{\partial s} +\frac{\partial}{\partial s}(\frac{4}{3}mu_i\frac{\partial v}{\partial s})
\end{equation}

\begin{equation}
    \frac{\partial E_e}{\partial t}+\frac{1}{A(s)}\frac{\partial}{\partial s}[(E_e+P_e)A(s)v]=\frac{1}{A(s)}\frac{\partial}{\partial s}(A(s)\kappa_{0e}T^{\frac{5}{2}}_e \frac{\partial T_e}{\partial s})+v\frac{\partial P_e}{\partial s}+\frac{k_B n}{\gamma - 1}\nu_{ie} (T_i - T_e)-R+H
\end{equation}

\begin{equation}
    \frac{\partial E_i}{\partial t}+\frac{1}{A(s)}\frac{\partial}{\partial s}[(E_i+P_i)A(s)v]=\frac{1}{A(s)}\frac{\partial}{\partial s}(A(s)\kappa_{0i}T^{\frac{5}{2}}_i \frac{\partial T_i}{\partial s})+v\frac{\partial P_i}{\partial s}+\frac{k_B n}{\gamma - 1}\nu_{ie} (T_e - T_i)+\frac{\partial}{\partial s}(\frac{4}{3}mu_iv\frac{\partial v}{\partial s})+\rho v g_{||}
\end{equation}

\begin{equation}
    E_e=\frac{P_e}{\gamma-1}, E_i=\frac{P_i}{\gamma-1}+\frac{1}{2}\rho v^2
\end{equation}

\begin{equation}
    P_{e,i} = 2k_BnT_{e,i}
\end{equation} where $\rho$ is the plasma density, $A(s)$ is the cross-sectional area, $s$ is the dimension along the loop, $v$ is velocity in the field-aligned direction, $g_{||}$ is the component of gravity along the field, $P$ is pressure, $E$ is energy, $\kappa$ is the conductivity, $T$ is temperature, $H$ is the heating function term, $R$ is the radiative loss function, $\nu_{ie}$ is the Coulomb collision frequency, $u_i$ is the ion viscosity, and $n$ is the particle density. HYDRAD solves the electron and ion temperatures separately, hence the subscripts on energy, pressure, and temperature above, and the presence of two energy equations; however, these deviate appreciably only in highly impulsive, low-density situations. For the simulations presented below, the deviations between the electron and ion temperatures remained negligible.

\subsection{Geometry \& Boundary Conditions}

To obtain the magnetic structure of our near-null loop we use the well-known embedded bipole field consisting of a uniform background vertical field and a vertically-oriented point dipole located some distance d below the photospheric surface. This produces an axisymmetric system and has been used by many authors for studies of coronal jets (e.g., \cite{Pariat2009,Wyper2018}). Note that the embedded bipole field is the simplest possible topology that exhibits a separatrix and is very frequently present in the actual corona. In cylindrical coordinates the field takes the simple form:

\begin{equation}
    \vec{B}_{total}=\vec{B}_0+\vec{B}_{dipole}
\end{equation}
where
\begin{equation}
    \vec{B}_{0}=B_0 \hat{z}
\end{equation}
and 
\begin{equation}
    \vec{B}_{dipole}=\frac{\mu_0 d}{4\pi}(\rho^2+(z+d)^2)^{-\frac{3}{2}}[3\rho(z+d)\hat{\rho}+(2(z+d)^2-\rho^2 )\hat{z}]
\end{equation}
The resulting field is shown in Figure \ref{F1}, exhibiting a null point along with a dome separatrix and spine lines, as expected. In the corona structures of this form have long been observed and identified as jets \citep{Bohlin1975}, anemones \citep{Shibata1994}, or fountains \citep{Tousey1973}, depending upon the scale size at which they form.  Several of the closed field lines under the fan surface are shown in Figure 1b. To define the loop geometry for the simulations presented here, we extracted the coordinates and arrays of $g_{||}$ and $B$ values of a single magnetic field line from a coronal jet simulation utilizing the field above \citep{Karpen2017}. 

\begin{figure}
\vspace{-\intextsep}
\centering
    \hspace*{-1\columnsep}
    \includegraphics[width=.5\linewidth,clip]{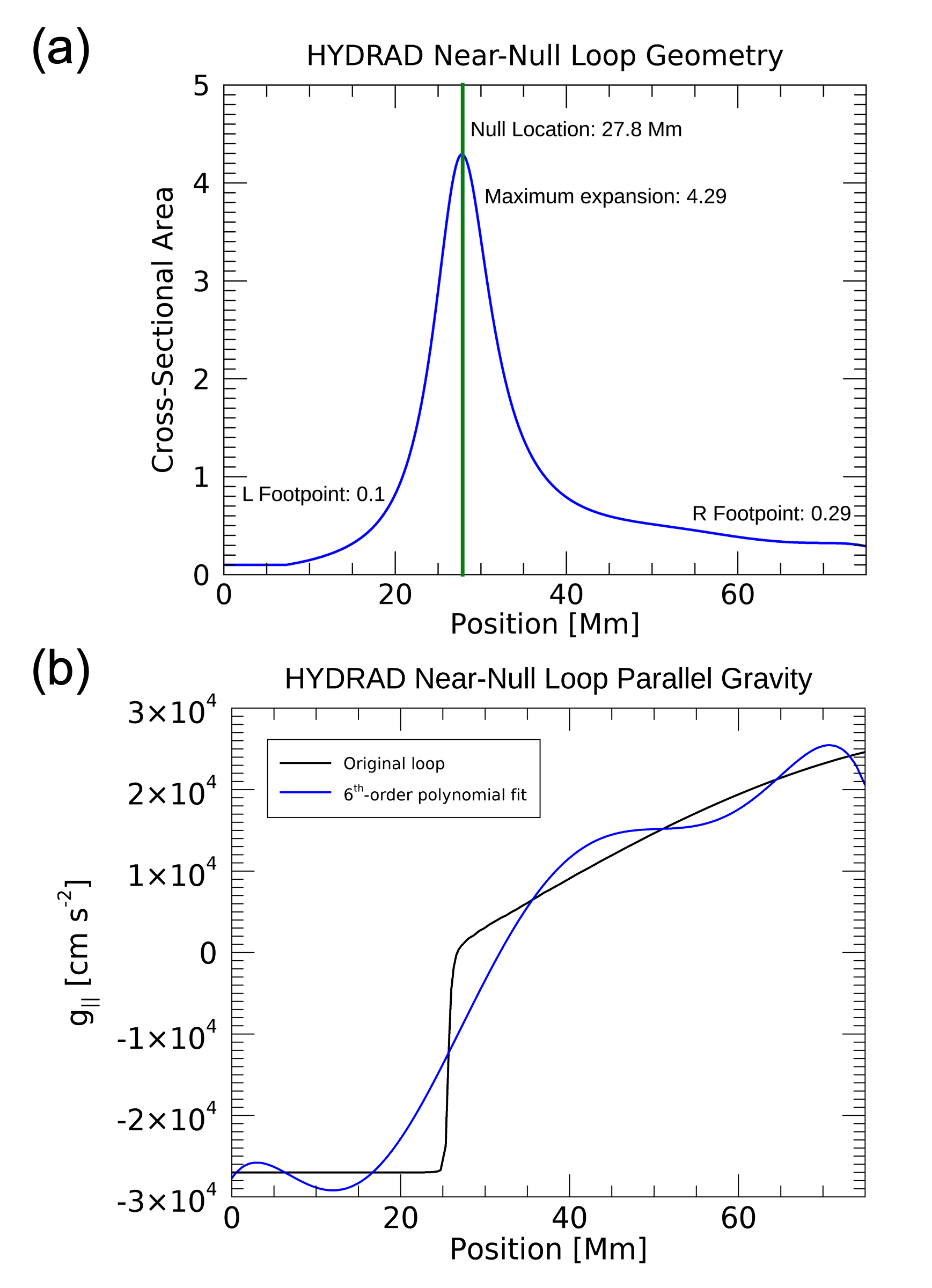}
              \caption{(a) graph showing the variation of the cross-sectional area along the length of the loop used in the HYDRAD code. The specific values at the footpoints and near the null are shown on the graph. (b) graph showing the variation of the parallel component of gravity along the length of the original loop (in black) and along the length of the loop used in the HYDRAD code (blue).}
              \label{F2}
\end{figure}

From the embedded bipole field above, and shown in Figure \ref{F1}, we selected a field line that passes very near to the coronal null seen in Figure 1b. Figure \ref{F2}a shows the cross-sectional area profile for the loop, and Figure \ref{F2}b shows the parallel gravity ($g_{||}$) for both the original loop and the polynomial fit which is the input for HYDRAD. The gravity profile is not significantly different than the coronal loops of the many previous studies, and while the polynomial fit is not as accurate as the magnetic field fit for the same loop, the relatively short length of the loop precludes significant impact to the simulation. The area factor, on the other hand, shows some important new features. First, we note that the left footpoint of the loop is located in a strong field region, while the right footpoint is located in field about one-third the strength and, consequently, the area on the right is approximately 3 times that on the left. More important, we note that there is a region of very strong expansion localized near $s = 28$ Mm. The maximum area there is roughly a factor of 40 greater than that at the left footpoint. We applied an $8^{th}$-order polynomial fit to the magnetic field to model the loop geometry in HYDRAD. The maximum standard fit in HYDRAD is $6^{th}$-order; however, this did not sufficiently capture the sharp apex of the loop, requiring us to increase the order. This again emphasizes the unique properties of near-null loops. The polynomial equations utilized for the loop in the simulations are

\begin{equation}
    B(s)=7802s^8-34,218s^7+ 62,460s^6-61,617s^5+ 35,866s^4-12,786s^3+2881s^2-417s+32.37
\end{equation}

\begin{equation}
    g_{||}=-(6.958\cdot10^6) s^6+(2.081\cdot10^7)s^5-(2.288\cdot10^7)s^4+(1.100\cdot10^7)s^3-(2.014\cdot10^6)s^2+(1.127\cdot10^5 )s-2.771\cdot10^4
\end{equation}
The maximum refinement level in an adaptive grid simulation determines the size of the smallest grid cell. For the results presented here, a maximum refinement level of 10 was selected, resulting in a minimum grid cell size of $1.22\cdot10^5$ cm. We additionally constrained the simulation to have a maximum grid cell width of $1.25 \cdot 10^8$ cm. Note that both footpoints are several scale heights deep in the chromosphere, so that there is essentially no evolution there.

\section{Simulation Results}\label{sec:results}

We executed two steady-state simulations, differentiated only by disparate background heating rates. The background heating rate in HYDRAD is determined by the initial conditions calculation, for which the user inputs the loop length, the distance between the tops of the constant-temperature chromospheric regions at each end (i.e., the coronal loop length), and the plasma density at these points. The top of the chromosphere, where the temperature gradient vanishes, is generally defined to be the loop ``base" in coronal loop models. The initial-conditions code then computes the steady background heating rate which would be necessary to sustain a symmetric, uniform cross-section loop in hydrostatic equilibrium with the given footpoint density, (see \citep{Bradshaw2004,Bradshaw2011a} for full details). We then apply the geometry seen in Figure \ref{F2} and allow the loop to find a new steady-state equilibrium, if one exists, with the previously-determined background heating.

For the simulation that we hereafter refer to as low-heat (LH), the code's standard base (top of the chromosphere) plasma density of $10^{11} \, cm^{-3}$ resulted in a background heating rate of $4.882\cdot10^{-4} \, erg\, s^{-1}\, cm^{-3}$. The high-heat (HH) run was set to a base density of $4\cdot10^{11}\, cm^{-3}$, resulting in a background heating rate of $2.417\cdot10^{-3} \,erg\,s^{-1}\, cm^{-3}$. Both of these values are typical for coronal loops.

\begin{figure}
\centering
    \includegraphics[width=.6\linewidth,clip]{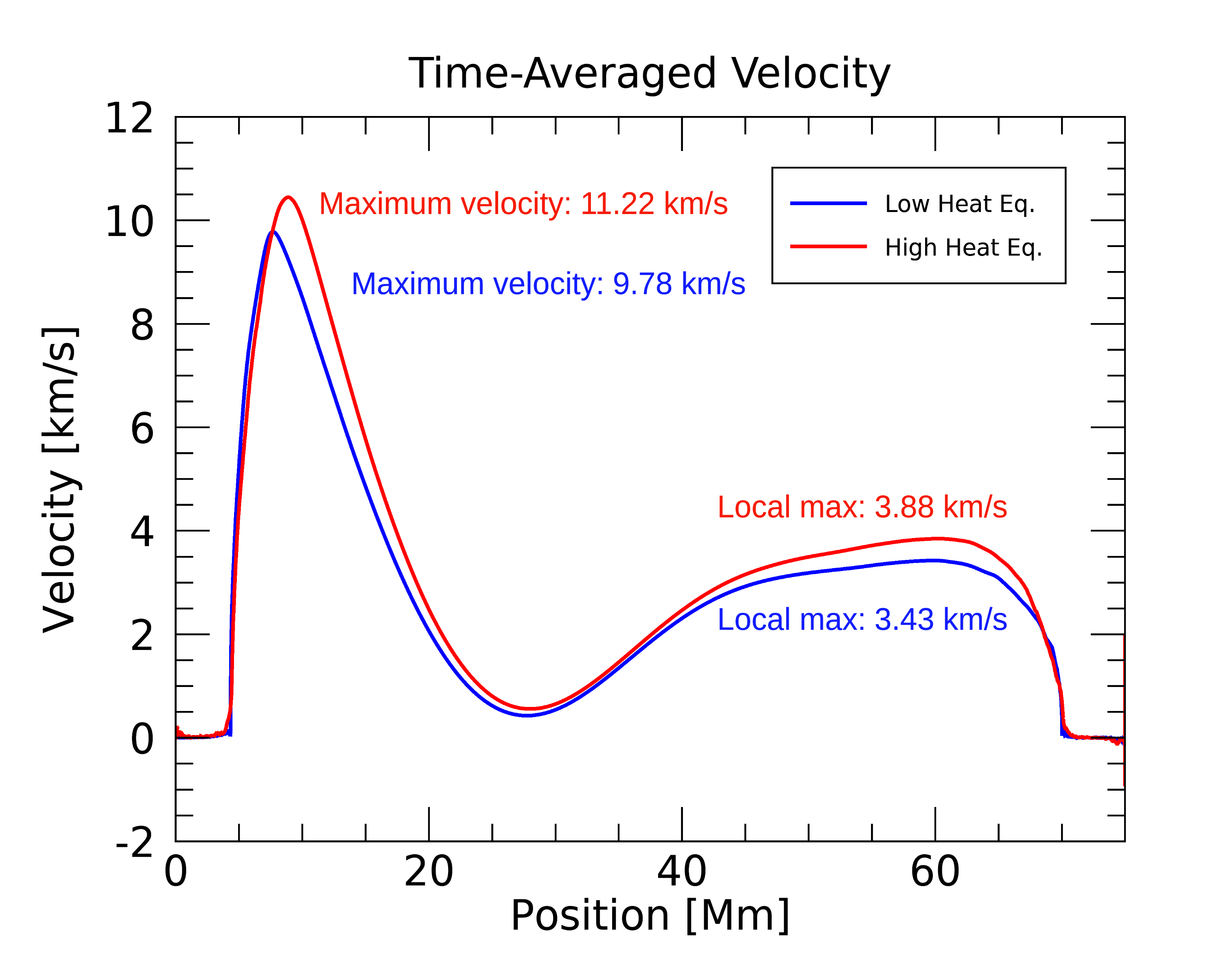}
              \caption{Time-averaged velocity graphs for LH (blue) and HH (red) steady state runs. The LH steady state velocity peaks on the left leg at 9.78 km s$^{-1}$, while the HH run produced a maximum velocity near the same point at 11.22 km s$^{-1}$. Both runs experienced near-null velocities of less than 1 km s$^{-1}$, and the right leg maxima were 3.43 and 3.88 km s$^{-1}$ for the LH and HH simulations, respectively.}
              \label{F3}
              
\end{figure}

Figure \ref{F3} shows the final velocities that we find from the two simulations, and is the primary result of our calculations. First, and most important, we find that these velocities are essentially steady. Both runs reached their steady state quickly after the beginning of the simulation; the LH steady state was established around 2.5 hours of elapsed solar time, while the HH converged even faster, after only 2.2 hours. We do find some ``jitter" in the solution due to numerical noise and waves bounding back and forth, but these time-average out to zero. Therefore, even for this extreme variation in the loop cross-section, a steady solution is possible -- at least for a near-uniform heating. The velocity profile, however, is distinctly different than that of previous loop models due to the unique nature of the area variation. 

The effect of a cross-sectional area variation $A(s)$ on loop plasma structure has been studied for many years \citep[e.g.,][]{Vesecky1979,Mikic2013}. In particular, recent studies by \citet{Klimchuk2019} and \citet{Cargill2022} present a thorough analysis on the effect of an area variation on loop plasma structure and dynamics, including detailed numerical simulations for spatially- and/or temporally-varying coronal heating. Here we only summarize the underlying physical understanding, but encourage the reader to refer to those papers for the rigorous modeling. Consider the classic loop model that has a uniform cross-section, $A(s) \equiv 1$, and is symmetric about the apex. Since the heat flux at the loop base vanishes, the total heating into the loop, $ E$, must be balanced by the radiation losses from the coronal and transition-region sections of the loop $E = R_{c} + R_{tr}$. It has been established from many simulations that the losses from these two regions are of the same order of magnitude, but with the transition region emission somewhat larger, $R_{tr} \approx 2 - 3 \times  R_c$ \citep[e.g.,][]{Klimchuk2008}. Note that for this loop the average area in the corona is equal to that in the transition region, since the area is constant everywhere. Now consider the effect of an area variation $A(s) \neq const$, as for example, in a typical dipole magnetic field, and assume that the area at the loop base $A_b = 0.1$ whereas that at the apex remains at $A_c = 1$. For this loop the area factor $\Gamma = 10$, which is a typical value.  The ratio of the radiation losses from the transition region versus that from the coronal section of the loop now changes due to the relative average area of these sections. Since the transition region lies at the very base of the loop and is extremely narrow, its average area $\langle A_{tr}\rangle \approx A_b$. As a result, the losses from this region are reduced by this factor, $R_{tr} \times A_b$, whereas the coronal losses and total heat input are reduced by the value of the area factor averaged along the whole loop $ \langle A(s) \rangle $. But $ \langle A(s) \rangle  > A_b $ so as a result, the loop is no longer in energy balance; the heating is larger than the radiative cooling, $E > R_{c} + R_{tr} \times \frac{A_b}{\langle A(s) \rangle}$. To establish a new balance the coronal losses must increase from their previous value, which requires that the density and hence the pressure increases. We conclude that an area constriction causes the coronal pressure to increase (see \citep{Klimchuk2019,Cargill2022} for quantitative details on how this increase varies with $A(s)$). If the area factor (effective constriction) is larger in the left leg than in the right leg, then the pressure will be larger on the left so that no static equilibrium is possible and a flow from left to right must occur. For example, if the left leg occurs in a high-field strength $B \approx 100 $ G active region, whereas the right leg is in the surrounding photosphere where typically, $B \approx 10$ G, then an area expansion of 10 will be present from left to right. This will result in a relatively-slow, $V < 10$ km/s, steady siphon flow \citep[e.g.][]{Antiochos84} that accelerates from the left leg, reaches a maximum in the corona and decelerates on the right.  

 These straightforward arguments describe the physical origin of the flows seen in Fig. \ref{F3}. Defining each loop leg from the apex, which coincides very closely with the point of maximum area, then the left leg has an effective $\Gamma$ factor much larger than that on the right, which accounts for the left-to-right flow. Such a flow would inevitably be present in any loop that lies very near a separatrix boundary, and would be relatively large compared to loops far from the separatrix, because the area variation is much larger at the separatrix, becoming near-singular there.

A unique feature of the flow is that it becomes very small near the apex (vicinity of null). The reason for this is straightforward. Since the loop is in a quasi-steady state, the flow must be steady so that the mass flux $nvA $ is constant. As in all coronal loop models, thermal conduction keeps the temperature in the coronal section of the loop near-constant; consequently, the density must also be nearly constant in the corona, otherwise there would be large pressure imbalances. The area, however, becomes very large near the null, so the velocity must become small there in order to maintain a near-constant mass flux. It should be emphasized that such a velocity profile would be a feature of only a separatrix loop, because it requires a very large and very localized expansion factor; in other words, a null point region. 

In terms of the quantitative values of the velocities for our particular parameters, the cases have different velocities, as one would expect, given the heating differential. There is a 1.5 km s$^{-1}$ difference in maximum velocity on the left leg (9.78 and 11.22 km s$^{-1}$ for LH and HH equilibria runs, respectively); the velocities near the null are similar, and both are approximately 0.5 km s$^-1$ (0.43 and 0.56 km s$^{-1}$), in essence, negligible velocity. On the right leg, the velocity picks up again, with the peaks at 3.43 and 3.88 km s$^{-1}$, the higher velocity belonging to the HH simulation. These values would be easily observable by spectroscopic instruments, and would be present in all coronal lines.

\begin{figure}
\centering
    \includegraphics[width=.6\linewidth,clip]{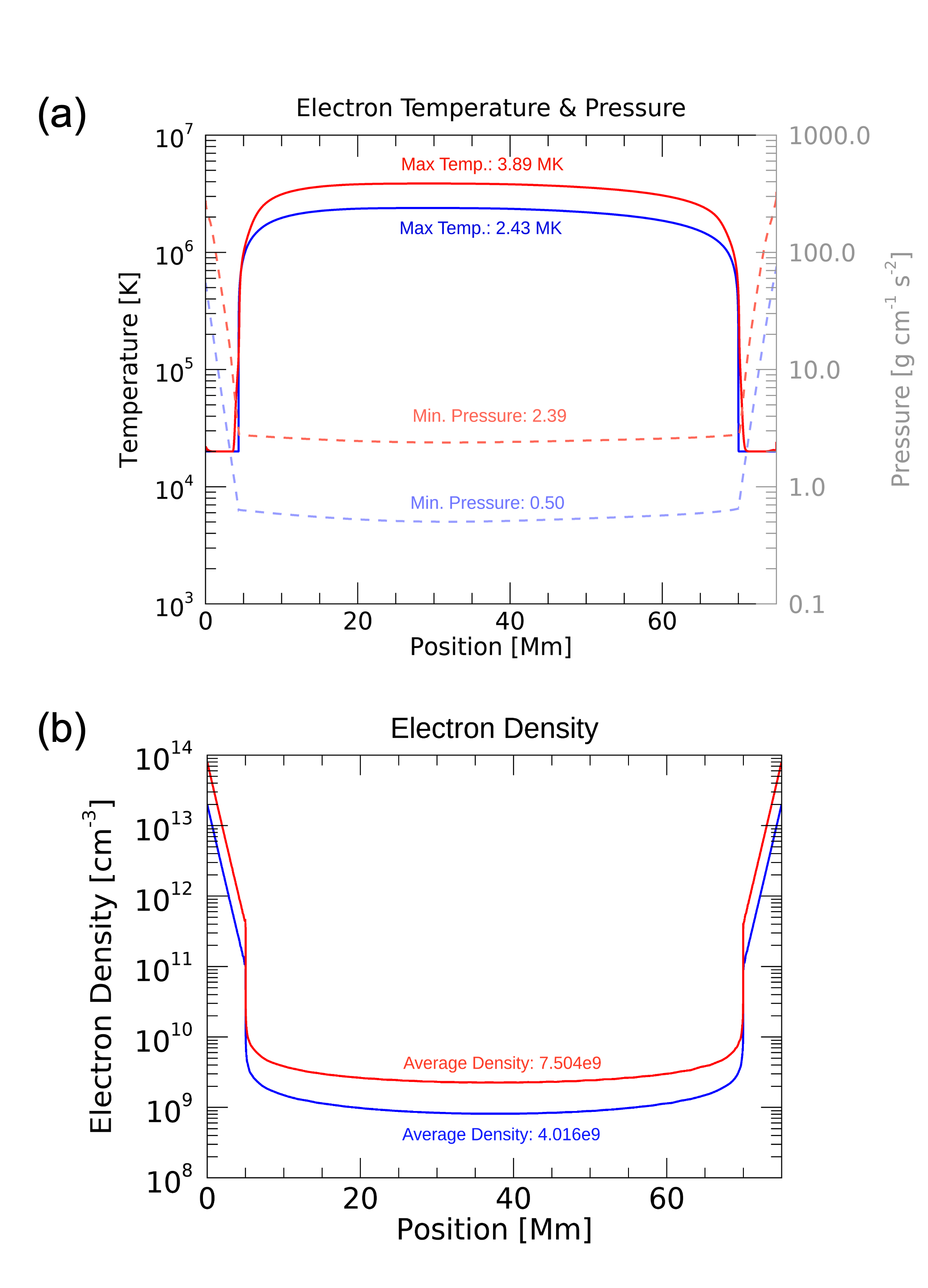}
              \caption{(a) Graph of the electron temperatures (solid lines) and pressures (dashed lines) for both the LH (blue) and HH (red) equilibria runs. The LH run reaches a maximum temperature of 2.43 MK, while the HH peaks at 3.89 MK. There is an over four-fold increase in electron pressure between the two runs, at 0.50 and 2.39 for the LH and HH runs. (b) Electron density for the two runs.}
              \label{F4}
\end{figure}

It should be emphasized that the flows in Figure \ref{F3} are all due to plasma pressure gradients; there is no other force in this 1D system other than gravity, which is always downward. Figure \ref{F4}a shows the electron temperatures and pressures for both runs. The peak loop temperature occurs to the left of the near-null region on both runs, and there is approximately a 1.5 MK difference between the temperatures (2.43 and 3.89 MK) in the two cases. Note that both cases show a small asymmetry in the density, with slightly higher density and a more gradual gradient on the right leg compared to the left (see Figure \ref{F4}b). These gradients are required to balance the flow profile, Fig. \ref{F3}, which shows a fast acceleration on the left, but a gradual deceleration on the right.  The pressure exhibits a rapid decrease on the left leg, which accelerates the flow upward, then bottoms out to a near constant level, except for the small variations required to balance gravity, and then increases on the right leg in order to decelerate the flow. Again, these gradients are set up by the need to maintain local thermal balance (heat in equals radiation out), given that the transition region radiation from the two legs is very different.

The HH run's minimum, coronal pressure (2.39 dyne cm$^{-2}$) is over 4 times higher than the LH run's (0.50 dyne cm$^{-2}$). This is quite close to what would be predicted by the standard scaling laws: $ P \propto H^{6/7}$ \citep[e.g.,][]{Rosner1978,Bradshaw19}. This result is to be expected given the arguments above. The scaling laws basically reflect the requirement that the coronal radiative losses must approximately balance the heating there, given that the transition region does not dominate the losses \citep{Rosner1978,Bradshaw19}. This requirement is even more valid in our case, because the transition region radiation is suppressed by the large expansion factors. Although there are strong flows in our model, they act only to redistribute the energy and are not an energy sink, because just like the conductive fluxes, they vanish at both footpoints. Note also that the flows become very small near the null region, so their enthalpy flux is negligible there. This has important implications for the thermal stability of the loop plasma, as will be discussed in Section \ref{sec:disc} below.

\section{\textit{Hinode} Observations}\label{sec:hinode}

The main conclusion from the results above is that coronal loops near separatrix surfaces should have distinctive velocity signatures. As a preliminary observational test of this hypothesis, we examined a well-observed and clearly identified null-point topology region for the type of flows predicted by our models. One caveat to our analysis -- and essentially to all observational inferences of coronal loop structure -- is that there appears to be a long-standing contradiction between the theoretical picture of loops as a representation of the variable magnetic field of a coronal flux tube, and observations that apparently show structures of nearly-constant cross-sectional area \citep{Klimchuk92,Peter12,Klimchuk20}. Most loop simulations, like this one, and many others \citep[e.g.,][]{Mikic13,Froment2018}, consider loops with varying cross-sectional area, as required by the conservation of magnetic flux. However, observations by some of the same authors, including \citet{Klimchuk20} and \citet{Auchere2018}, consistently show loops with cross-sectional areas that vary very little, even in the vicinity of higher-beta regions. An alternative approach to apparent loop geometries is presented by \citet{Malanushenko2021}, but that paper focuses mostly on more traditional active region loops where the low-beta approximation is valid. As has been discussed in \citet{Mason2019}, which first reported the NPTs presented here, there are frequently distinct dark regions around the null of such structures. It is likely that these are caused by highly-expanded loop structures like the one that we model here. It is difficult, however, to understand how a loop that appears to have constant cross-section could sustain the pressure differential that would attend the required variance in plasma pressure along its length, particularly with a flow present. Without reliable estimates of the magnetic field strength in these low-coronal regions, the mystery persists. In order to resolve the loop-area contradiction, it is clear that much more work is required, both modeling with a 3D MHD code with full thermodynamics as in  \citet{Mikic2018}, and direct spectropolarimetric measurements of the magnetic fields in and around null-point structures as, perhaps, with the Daniel K. Inouye Solar Telescope.

Ignoring the contradiction above, we now present observations of several NPTs imaged by the Hinode EIS instrument, to investigate the evidence for long-lived end-to-end flows \citep{Klimchuk2019} and thereby test our conclusions above.  The two structures shown here have already been investigated in detail in \citet{Mason2019} and \citet{Mason2021}, and the reader is referred to those works for details on their evolution and dynamics; here we simply state that they are NPTs formed from decaying active regions which form on the border of (or entirely within) coronal holes.

\begin{figure}
    \centering
    \includegraphics[width=.65\linewidth,clip]{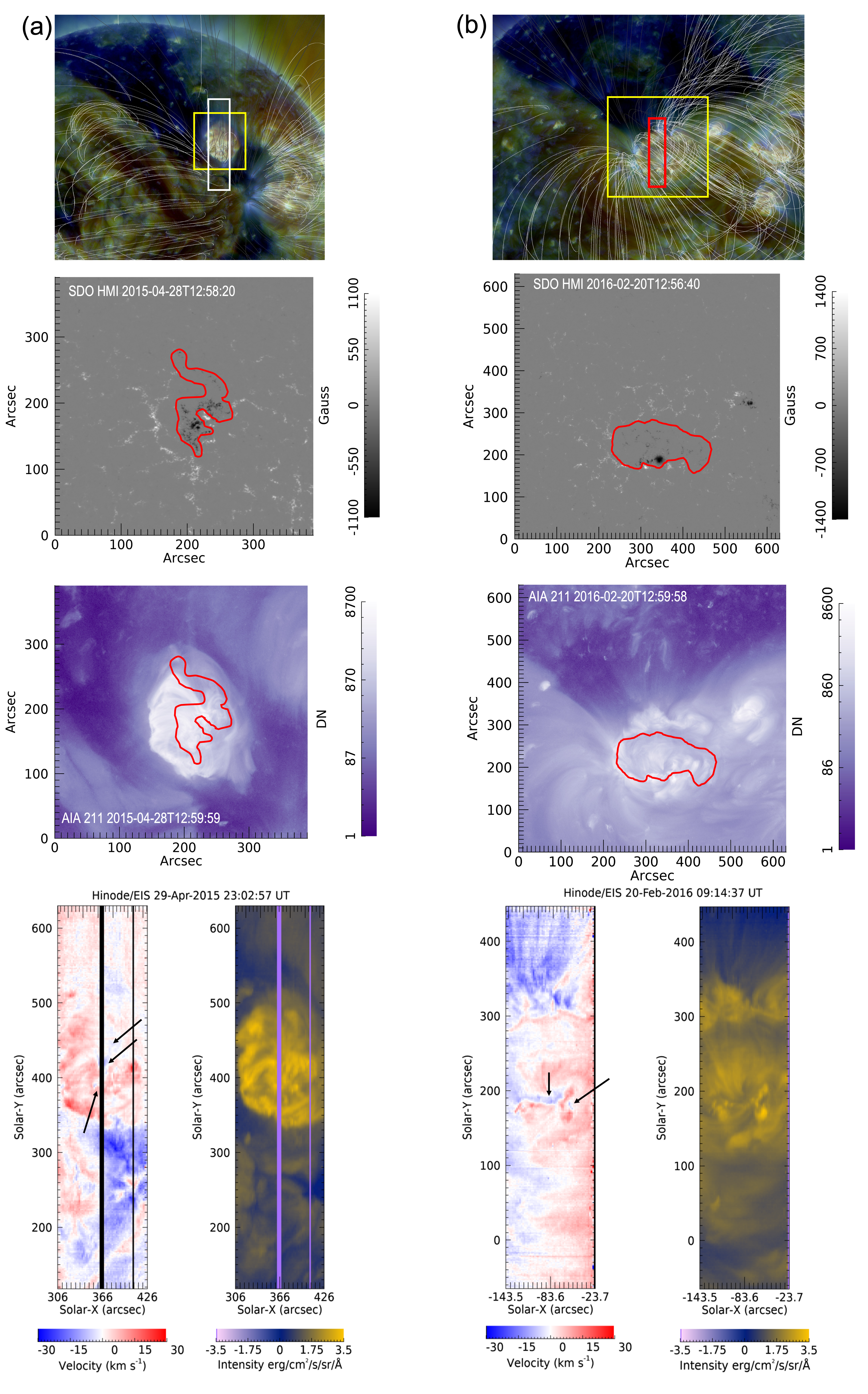}
              \caption{(a) This column presents the null-point topology observed by Hinode EIS on 2015 April 29. The top image is a 171-193-211 Å composite AIA image for context, overlaid with a potential field source surface, from The Sun Today. The white rectangle shows the FOV for the EIS observation at the bottom, and the yellow square shows the FOV of the SDO HMI and AIA cutouts immediately below. The AIA and HMI cutouts are overlaid with the approximate location of the polarity inversion line within the NPT. At the bottom are the Hinode EIS observations using the 195.12 Å wavelength. Closed loops can be seen in the intensity map, running from several inner spines along the long axis of the ovate dome to the outer fan surface. The velocity map shows several locations, marked with black arrows, where the flows are upward from the several inner spines; the entire outline of the fan surface, however, is marked with strong downflows. (b) Analogous observation of an NPT on 2016 February 20; locations of upflows are again marked with black arrows on the velocity map, and velocities are similar in range.}
              \label{F5}
\end{figure}

Figure \ref{F5} shows two NPTs as seen in the Fe XII 195.12 Å wavelength; the peak formation temperature for this line is 1.58 MK. In both \ref{F5}a and \ref{F5}b, there are distinct regions around the inner spine of the NPTs that have upflows on the order of 3-6 km s$^{-1}$. Note that for a complex polarity distribution such as found on the true photosphere -- especially those that give rise to so-called pseudostreamers -- the ``spine" often corresponds to a sheet-like structure \citep[e.g.][]{Titov2011,Scott21}. On the opposite end of these closed loops, around the fan surface, there are strong downflow signatures, which peak between 10 (Figure \ref{F5}b) and 16 km s$^{-1}$ (Figure \ref{F5}a). This general pattern correlates with the simulation; the positive flows in the direction of the loop on both legs correspond to Doppler shifts in the observations that show upflows on the spine leg and downflows on the fan leg. Interestingly, the magnitude of the upflows on the spine legs in the observations is consistently lower than the downflows on the fan leg, in contradiction to the simulation results. This could be due to observational affects, such as the inner spine being at an angle from the perpendicular with respect to the instrument, thereby lowering the vertical component of the upflow. A key point is that overlying coronal loops that have a strong overall net flow could weaken the upflow and strengthen the downflow signatures.  Both NPTs discussed here are adjacent to a helmet streamer, and it has been argued recently that such flows may be common there due to the effects  of thermal nonequilibrium \citep{Schlenker21}. Alternatively, the dynamics of the near-null expansion in creating the strong upflow along the spine leg could be somewhat less exaggerated on the Sun due to effects such as field-line tangling or impulsive heating that are not included in the calculations above. Further study is required to determine other factors that might be affecting the distinction in velocities between the two legs, and whether this is a consistent discrepancy across NPT closed-field loops. The basic flow pattern, however, agrees very well with our loop modeling and demonstrates the distinctive nature of flows near separatrices.

\section{Discussion}\label{sec:disc}

The results above have a number of important implications for understanding the structure and dynamics of coronal plasma. One key conclusion is that flux tubes (coronal loops) near a separatrix in the closed corona should exhibit strong persistent flows, with a direction determined by the topology of the separatrix, spine vs fan. One caveat to this conclusion that merits some discussion is the validity of the 1D model for near-null loops. The underlying assumption in all coronal loop models is that the plasma beta is so low that the field can be assumed to be rigid throughout the plasma evolution. But if a loop has a very large expansion at some location, it implies that the field strength becomes small there, so the low beta assumption becomes suspect. In our loop, for example, the field strength is down by a factor of roughly 40 from its value at the chromospheric base, so the field energy density decreases by almost three orders of magnitude over the same distance. We conclude that simulations of observed loops near separatrices should be checked \textit{a posteriori} to verify that the low-beta assumption does remain valid. If the beta does become significantly larger than unity, the plasma pressure is expected to cause the field near the null to expand even further, thereby enhancing the flows found above. It is also possible that such flows induce interchange reconnection, which would induce additional dynamics as found by \citet{Scott21}, for example. Furthermore, magnetic field changes could well lead to non-steady flows and oscillations in the field. However, the thermal conduction and mass flows will still be along the field-aligned direction, which may maintain similar dynamics to truly low-beta regions. We conclude, therefore, that MHD modeling of near-null loops would be highly informative, and plan to undertake such a study in a subsequent paper.

For field-aligned models, our work proves the viability of an end-to-end flow as a stable steady state solution in coronal loops with extreme geometry. We find that such a flow establishes itself rapidly as a stable steady state for a range of heating values in a loop that is highly asymmetric both in its apex location and cross-sectional area variation. The flow is from left to right, as expected from simple considerations of loops with asymmetric expansion factor. It is well known that if everything else is kept constant, increasing the expansion factor increases the coronal pressure, because the relative amount of transition region emission decreases \citep{Klimchuk2008,Cargill2021}. Since the left leg of our loop has a larger corona-to-footpoint expansion factor, it has a larger pressure, which drives a flow from left to right.  This finding supports the many observations of flows in coronal loops in structures like active regions, observed even when there is no apparent time-dependent heating occurring. Previous work has suggested that such apparent motion may be due to waves propagating along the loop instead of actual plasma motion, but this work shows that stable end-to-end flows can be maintained in a broad range of loop geometries and situations.

Although we do find stable steady state loops, it should be emphasized that they are likely to be more susceptible to thermal instability than standard loops. The key point is that in the vicinity of the null, the energy balance is primarily between heating and radiative losses. As discussed above, the transition region losses are highly suppressed by the large expansion factors. Also, the large expansion causes the flows and, hence, enthalpy flux to be small there. If the thermal and enthalpy fluxes were completely negligible, so that the energy balance were only between a constant heating and coronal radiation, then the plasma there would be more likely to be thermally unstable \citep[e.g.][]{Parker53,Field65}. We conclude that the small residual fluxes are providing the stability in our solutions. The implication, however, is that as one considers loops that are closer and closer to the separatrix and, therefore, have an increasingly larger localized expansion, then the plasma there would become thermally unstable. In fact, we have observed coronal rain that we attributed to thermal non-equilibrium preferentially near the separatrix surfaces of null-point topologies \citep{Mason2019}. We intend to investigate this issue in detail with further modeling work. 

The most important result from our work is that \textit{the topology of the coronal magnetic field, in particular the imprint of the S-Web at the top of the corona, should be observable with high-resolution spectroscopic measurements}. Although we have used the simplest possible coronal heating, constant in space and time, our results must hold in general: fast flows (exceeding 10 km s$^{-1}$) will be induced by the extreme flux tube geometry at separatrix boundaries. In fact, the data presented in Section \ref{sec:hinode} appear to support this conclusion. Note that as shown above, the magnitude of the steady flows is sensitive to the magnitude of the heating and the loop geometry.  Furthermore, the flows will be sensitive to any spatial variation in the heating, for example, if the strong field region on the left has higher heating than the right. Precise measurements of the plasma velocities would provide stringent constraints on such variations. Note also that although our single-null-point loop is likely to be the most common separatrix structure, it is not unlikely to have flux tubes that pass by two null points as in models of large pseudostreamers \citep[e.g.][]{Titov2011,Scott21}. Such loops would have multiple localized high-expansion regions with even more complex dynamics than our case above, and may well be even more sensitive to the details of the heating. The wave heating models, in particular, are sensitive to the magnetic field variations along the loop, because the heating depends strongly on wave reflection from variations in the Alfvén speed along the loop \citep[e.g.,][]{vanderHolst14,Downs16}. Near-null loops would have extreme variations in the Alfvén speed. Our results and possible extensions, therefore, constitute a potentially powerful observational probe of not only coronal magnetic topology, but perhaps even of the heating mechanism itself. It is clear that a great deal more theoretical and observational work remains to be done on this important topic of coronal loops near separatrices.

\section{Acknowledgements}

EIM’s research during the development of this paper was supported by an appointment to the NASA Postdoctoral Program at the NASA Goddard Space Flight Center, administered by Universities Space Research Association under contract with NASA. EIM would like to acknowledge support from the NASA LWS program. SKA was supported by LWS Strategic Capability Grant 80NSSC22K0892 to the University of Michigan.

\bibliography{HYDRADeq}{HYDRADeq}
\bibliographystyle{aasjournal}



\end{document}